\documentclass[twocolumn,secnumarabic,amssymb, nobibnotes, aps, prx, superscriptaddress, nobalancelastpage,longbibliography]{revtex4-2}

 \usepackage{graphicx}
\usepackage{bm}
\usepackage{amsmath} 
\usepackage{braket} 
\usepackage{epsfig}
\usepackage{tensor}
\usepackage{CJKutf8}
\usepackage[version=4]{mhchem}
\usepackage{titlesec}
\usepackage{ifthen}
\usepackage{sidecap}
\usepackage{listings}
\usepackage{array}
\usepackage{floatrow}
\usepackage{enumitem}
\usepackage[para,online,flushleft]{threeparttablex}
\usepackage{float}
\floatstyle{plaintop}
\restylefloat{table}

\usepackage{hyperref}
\hypersetup{breaklinks=true,colorlinks=true,linkcolor=blue,citecolor=blue,filecolor=magenta,urlcolor=cyan}

\usepackage[all]{hypcap}

\usepackage{xcolor}
\definecolor{pastelgray}{rgb}{0.81, 0.81, 0.77}
\definecolor{beaublue}{rgb}{0.9, 0.9, 0.93}

\usepackage{tikz,xcolor,hyperref}

\definecolor{lime}{HTML}{A6CE39}
\DeclareRobustCommand{\orcidicon}{
	\begin{tikzpicture}
	\draw[lime, fill=lime] (0,0) 
	circle [radius=0.16] 
	node[white] {{\fontfamily{qag}\selectfont \tiny ID}};
	\draw[white, fill=white] (-0.0625,0.095) 
	circle [radius=0.007];
	\end{tikzpicture}
	\hspace{-2mm}
}

\foreach \x in {A, ..., Z}{%
	\expandafter\xdef\csname orcid\x\endcsname{\noexpand\href{https://orcid.org/\csname orcidauthor\x\endcsname}{\noexpand\orcidicon}}
}

\makeatletter
\def\@bibdataout@aps{%
\immediate\write\@bibdataout{%
@CONTROL{%
apsrev41Control%
\longbibliography@sw{%
    ,author="08",editor="1",pages="1",title="0",year="1"%
    }{%
    ,author="08",editor="1",pages="1",title="",year="1"%
    }%
  }%
}%
\if@filesw \immediate \write \@auxout {\string \citation {apsrev41Control}}\fi
}
\makeatother

\renewcommand{\vec}[1]{\mbox{\boldmath $#1$}}

\newcolumntype{Y}{>{\centering\arraybackslash}X}

\begin{document}
\begin{CJK*}{UTF8}{gbsn}

\title{Probing the Non-exponential Decay Regime in Open Quantum Systems}

\author{S. M. Wang (王思敏)\orcidA{}}
\affiliation{Key Laboratory of Nuclear Physics and Ion-beam Application (MOE), Institute of Modern Physics, Fudan University, Shanghai 200433, China}
\affiliation{Shanghai Research Center for Theoretical Nuclear Physics,
NSFC and Fudan University, Shanghai 200438, China}

\author{W. Nazarewicz\orcidB{}}
\affiliation{Facility for Rare Isotope Beams and Department of Physics and Astronomy, Michigan State University, East Lansing, Michigan 48824, USA}

\author{A. Volya\orcidC{}}
\affiliation{Department of Physics, Florida State University, Tallahassee, Florida 32306, USA}

\author{Y. G. Ma (马余刚)\orcidD{}}
\affiliation{Key Laboratory of Nuclear Physics and Ion-beam Application (MOE), Institute of Modern Physics, Fudan University, Shanghai 200433, China}
\affiliation{Shanghai Research Center for Theoretical Nuclear Physics,
NSFC and Fudan University, Shanghai 200438, China}

\begin{abstract}
The most important law of radioactivity is that of the exponential decay. In the realm of quantum mechanics, however, this decay law is neither rigorous nor fundamental. The deviations from the exponential decay  have been observed experimentally at the early stage of a decay process, but there is little evidence for non-exponential behavior 
at long times. Yet such long-term non-exponentiality is expected theoretically to probe the non-resonant background components of 
the initial wave function which preserve the  structural interference and the memory of how the state was created. In this paper, we propose new observables  that can be used for experimental investigations of the post-exponential decay regime, including the decay of threshold resonances, particle correlations in  three-body decays, and interference between near-lying resonances.   While the specific examples presented in this work pertain to atomic nuclei, the  properties of non-exponential decay are generic, i.e., they apply to other many-body open quantum systems, such as hadrons, atoms, molecules, and nanostructures. 
\end{abstract}

\date{\today}

\maketitle
\end{CJK*}

\section{Introduction}

The classical concept of radioactive decay comes from the  understanding that the rate of decay must be proportional to the amount of available material. An important assumption behind this is that decay is
 a stochastic process at the level of individual decaying particles; this implies that
the probability of decay does not depend on the previous history. The first quantum theory of radioactive decays has been developed in the late 1920s\,\cite{Gamow1928,Gurney1929,Weisskopf1930}. Since then, it became evident   that the  exponential decay law  {\it is the result of somewhat delicate approximations}\,\cite{Merzbacher1998} and cannot be exact\,\cite{Hellund1953,Khalfin1958,Levy1959,Schwinger1960,Winter1961,Newton1961,Goldberger1964a,Goldberger1964a,Baz1969,Fonda1978,Greenland1988,Esposito2004}.

At the initial time corresponding to  the formation of radioactive state, the decay rate must vanish\,\cite{Fonda1978,Peshkin2014}. The resulting initial-stage non-exponentiality has been verified experimentally\,\cite{Wilkinson1997}. The early-time dynamics is closely tied to questions of memory effect, quantum state preparation, eigenstate thermalization\,\cite{Fonda1978,Gorin2006,Deutsch2018,Volya2020art}, measurement theory, and Zeno paradox\,\cite{Ghirardi1979,Chiu1977,Kofman1996}. The behavior of decaying systems at long times has also been discussed extensively, see, e.g.,  Refs.\,\cite{Nicolaides1996,Mercouris1997,Joichi1998,Torrontegui2010,Garcia-calderon2019}. In spite of compelling theoretical arguments for decay's non-exponentiality  at long times, the experimental evidence for this phenomenon  is still missing {\it for isolated resonances}\,\cite{Nikolaev1969,Norman1988,Kelkar2004}. 
One  confirmed case of turnover into the non-exponential  decay regime comes from the study
of organic molecules interacting with the solvent environment which results in energy
 broadening  \cite{Rothe2006}. For such cases, the broadening of the exponent of multiple sources results in a $1/t^{2}$ distribution at long times \cite{Newton1961}.

Following the idea that the properties of an open quantum system can be influenced by its environment\,\cite{Torrontegui2010,Lawrence2002,Parrott2002}, the non-exponential decay has been investigated in cosmology\,\cite{Krauss2008,Urbanowski2011}, nanocrystals\,\cite{Sher2008}, and quantum dots\,\cite{Andersson2019,Yoshimi2022}. However, the direct measurement of the post-exponential decay for a single  metastable state still represents an appreciable challenge. This is because  a large amount of radioactive material is needed and/or a long observation period is required for such tests, and also because  environmental decoherence effects  are expected \cite{Beau2017,Greenland1988}.

In this paper, we discuss the properties of the quantum decay at long times, with a focus on the low-energy behavior and observables that  
shed light on the transition to the post-exponential regime. 
Specifically, we discuss such observables  in three dripline isotopes: (i) the newly discovered proton-rich $^9$N, in which broad threshold resonances are expected \cite{Charity2022}; (ii) the two-proton ($2p$) emitter $^6$Be \cite{Egorova2012}, illustrating the general case of the  three-body decay; and (iii) an artificially unbound neutron-rich $^6$He, called 
$^6$He$^{\prime}$. We want to emphasize that the purpose of this work is to explore the universal properties of post-exponential decay in dripline nuclei, how such effects could be observed, and what can generally be learned about open quantum systems from these observations. For this reason, we do not make detailed predictions for concrete nuclei. Still, we suggest promising candidates for experimental searches.

\section{Method}

\subsection{Hamiltonian and wave function}

To capture the long-time behavior of a decaying system, we utilize a recently-developed time-dependent approach\,\cite{Wang2021}. 
The intrinsic Hamiltonian of the $N$-body system of constituent clusters of masses $m_i$ and momenta $\vec{p}_i$ can be written as
\begin{equation}\label{Hcnn}
	\hat{H} = \sum_i^N\frac{ \hat{\vec{p}}^2_i}{2 m_i} +\sum_{i<j}^N \hat{V}_{ij}(\vec{r}_{ij}) -\hat{ T}_{\rm c.m.},
\end{equation}
where $\hat{V}_{ij}$ represents the pairwise interaction between the constituents as a function of relative distance $\vec{r}_{ij}$; and $\hat{T}_{\rm c.m.}$ stands for the kinetic energy of the center-of-mass. In practice, relative Jacobi  coordinates are used since they are more appropriate for describing the asymptotic behavior of the decaying system.

The initial wave function  $\Psi(0)$ is propagated using the time evolution operator $\exp({-i\hat{H}t/\hbar})$.
The exact form of the initial real-energy wave packet $\Psi(0)$ is important for our studies. Any such packet has a resonant part associated with the complex-energy Gamow pole(s) nearby, as well as a non-resonant scattering component. The former exhibits the usual exponential decay, while the latter is responsible for  the non-exponential decay at remote times \cite{Peshkin2014,Ramrez2019,Ramrez2021}. In the complex-energy framework, the Gamow state has  the purely outgoing wave asymptotics, and has a fixed structure that does not carry any dynamical information related to its formation.  It is associated with complex energy $\tilde{E} = E_r - i\Gamma/2$  where $\Gamma$ is the  decay width that defines 
the  half-life.  To generate the initial state, we start with 
a complex-energy resonant state $\tilde{\Psi}(0)$  
generated by the Gamow coupled-channel (GCC) method\,\cite{Wang2019,Wang2022}, in which the Schr{\"o}dinger equation is solved by utilizing the Berggren ensemble\,\cite{Berggren1968,Michel2009,Wang2017} in the complex-momentum $k$-space. 
This wave function is then projected onto the real-energy initial state $\tilde{\Psi}(0)\rightarrow \Psi(0)$ by expanding $\tilde{\Psi}(0)$ in real-momentum space through a Fourier–Bessel transformation within a 15\,fm spherical box\,\cite{Wang2022}. 

\subsection{Spectral function}

The Gamow resonant state is the eigenstate in the complex-energy plane. In the Hilbert space, this state become a wave packet whose wave function $|\Psi\rangle$  can be expanded in the basis of   scattering states \cite{Wang2021,Ramrez2021}, which are the real-energy eigenstates $|E\rangle$'s of the Hermitian Hamiltonian above the decay threshold. Due to the many-body correlations, both the resonant and scattering states contain multiple configuration components. Unlike the pure outgoing boundary condition that the resonant state  obeys, the scattering states  are described by  both incoming and outgoing  probability current in the asymptotic region \cite{Michel2021}. Consequently, for a given $E$, a system/Hamiltonian can have multiple degenerated scattering states. Since the nuclear interaction is short ranged, the asymptotic region can be decoupled and dominated by the kinetic term of the Hamiltonian (and long-ranged Coulomb for the charged particles). Through the unitary transformation,  the degenerate scattering states can be rearranged to make each governed by asymptotic configuration $c_i$ \cite{Fossez2013}. 
Consequently, the resonance wave packet $\Psi(t)$ can be written as 
\begin{equation}\label{Wave_function}
    \Psi(t) = \int \sum_i a_i(E)~e^{-i\frac{E}{\hbar}t}~| E, c_i \rangle ~{\rm d}E,
\end{equation}
where $a_i(E)$ represents the amplitude of each asymptotic configuration $c_i$, and the corresponding weight is $\mathcal{W}_i(E) = |a_i(E)|^2$. In practice, $a_i(E)$ can be obtained by analyzing the asymptotic wave function or the decaying products during the time propagation.

The survival amplitude $\mathcal{A}(t)$  is defined as the overlap between the initial state $\Psi(0)$ and the propagated state $\Psi(t)$ at time $t$. In our work, $\mathcal{A}(t)$ is evaluated  through the Fourier transformation of the spectral function $\rho(E)$  \cite{Fock1947}:
\begin{equation}\label{Survival_probability_expression}
	\mathcal{A}(t) = \braket{\Psi(0) | \Psi(t)} = \int_0^{+\infty} \rho(E) e^{-i\frac{E}{\hbar}t} {\rm d}E,
\end{equation}
where $\rho(E) = |\langle E | \Psi\rangle|^2$ can be obtained by expanding $\Psi$ in the real-energy eigenstates, or analyzing the asymptotic wave function at long times. 
Both ways are used to cross-check the results. The survival probability is directly obtained from $\mathcal{A}(t)$:  $\mathcal{S}(t) = |\mathcal{A}(t)|^2$.

The spectral function $\rho(E)$ links the non-exponential decay to the broad  near-threshold structures. Based on Eq.\,(\ref{Wave_function}), it can be written as
\begin{equation}\label{Spectral_function}
\begin{aligned}
     \rho(E) & =\sum_i |\langle E, c_i | \Psi(t) \rangle|^2,   \\
     & = \sum_i \mathcal{W}_i(E).
\end{aligned}
\end{equation}
Since the time evolution operator only changes the phase of the real-energy eigenstate $|E\rangle$, $\rho$ does not depend on time $t$.

Alternatively, the spectral function $\rho$ can be obtained by the energy derivative of the scattering phase shift ${\delta}_\ell$ \cite{Kelkar2004}, i.e., the level density \cite{Beth1937}. Due to   the centrifugal barrier, the near-threshold behavior of $\rho$ depends on the orbital angular momentum $\ell$ in the two-body system. By making use of the Mittag-Leffler expansion, in the absence of Coulomb interaction, the near-threshold spectral function becomes\,\cite{Ramrez2018}
\begin{equation}\label{rho3B}
\rho(E) = \frac{2\ell+1}{\pi}\frac{{\rm d} \delta_\ell}{{\rm d} E} =\operatorname{Im} \sum_n \frac{2\ell+1}{\pi\tilde{E}_{n}^{ \ell+1/2}} \frac{E^{\ell+1 / 2}}{\tilde{E}_{n}-E},
\end{equation}
where $\tilde{E}_{n}$ is the complex energy of the $S$-matrix pole.
In the three-body framework, the kinetic operator is\,\cite{Raynal1970,Wang2017}
\begin{equation}\label{EkinHH}
\hat{T} =-\frac{\hbar^2}{2 m}\left(\frac{1}{\mathcal{R}^{5/2}} \frac{\partial^2}{\partial \mathcal{R}^2}\mathcal{R}^{5/2}-\frac{\hat{K}^2+15/4}{\mathcal{R}^2}\right),
\end{equation}
where $\mathcal{R}$ is the hyperradius. The eigenvalues of the five-dimensional angular momentum operator
$\hat{K}^2+15/4$  are ($K$+3/2)($K$+5/2), and $K$ is the hyperspherical quantum number. Noticing that the
second term in Eq.~(\ref{EkinHH}) represents the centrifugal barrier, one can replace $\ell$ with $K+3/2$ in Eq.~(\ref{rho3B})
\cite{Ramrez2018}. This yields 
$\rho(E) \propto E^{K+2}$ at near-threshold energies for the charge-less systems. The strong channel dependence of $\rho(E)$ gives rise to  deviations from the Breit-Wigner form factor at low energies.

\subsection{Nucleon-nucleon correlations}

In  three-body (two-nucleon) decay experiments, the energy and angular correlations of the emitted particles can be directly measured. Since the nucleon-nucleon correlation strongly depends on the structural information, especially the asymptotic configurations shown in spectral function $\rho$, it is impacted by the presence of the non-exponential component in $\mathcal{A}(t)$. Expressing the energy correlation $\mathcal{C}$ in Jacobi-T coordinates, it can be written in terms of solutions propagated for long times:
\begin{equation}\label{Energy_correlation_general}
     \mathcal{C}(E_{pp}, E) =\lim_{t\rightarrow \infty}  \langle \Psi(t) | \delta(\epsilon - \frac{E_{pp}}{E}) | \Psi(t) \rangle, 
\end{equation}
where $E$ corresponds to the decay energy $Q_{2p}$ for the $2p$ decay and  $\epsilon$ is the ratio between the kinetic energy of the relative motion of the emitted nucleons and $E$. Utilizing Eq.\,(\ref{Wave_function}) and the fact that $\epsilon$ only depends on the hyperspherical harmonics in momentum space\,\cite{Raynal1970}, $\mathcal{C}(E_{pp}, E)$ becomes
\begin{widetext}
\begin{equation}\label{Energy_correlation}
\begin{aligned}
     \mathcal{C}(E_{pp}, E) & = \sum_{i, j} a^*_i(E) a_j(E) \langle c_i | \delta(\epsilon - \frac{E_{pp}}{E}) | c_j \rangle,   \\
     & = \sum_i \mathcal{W}_i(E) \langle c_i | \delta(\epsilon - \frac{E_{pp}}{E}) | c_i \rangle + \sum_{i \neq j} a^*_i(E) a_j(E) \langle c_i | \delta(\epsilon - \frac{E_{pp}}{E}) | c_j \rangle.
\end{aligned}
\end{equation}
\end{widetext}
As shown below, for each configuration $c_i$, the diagonal part $\langle c_i | \delta(\epsilon - {E_{pp}}/{E}) | c_i \rangle$ shows a distinct energy correlations as quantum numbers $(K,\ell_x,\ell_y)$ change. Similar to the spectral function, the energy dependence of the relative weight $\mathcal{W}_i/\sum \mathcal{W}_i$ or amplitude $c_i$ will also manifest itself in the asymptotic correlation, and this can be used to  quantify the presence of the  non-exponential decay component.

\subsection{Model space and parameters}

The extremely proton-rich $^9$N was recently discovered to have a $1/2^+$ ground state with a one-proton decay energy $Q_p=2.5$\,MeV and a width $\Gamma=1.8$\,MeV\cite{Charity2022}. This state is  the analog of the 1/2$^+$ ground state of $^9$He, which is also a good candidate for observing non-exponential decay. Additionally, the first excited state of $^9$He has a spin-parity $J^\pi$ = 1/2$^-$, and the measured one-neutron decay energy is $Q_n=2.2$\,MeV with a width $\Gamma=0.1$\,MeV \cite{ENSDF}. 
Due to the large spectroscopic factor\,\cite{Charity2022},  the nucleus $^9$N ($^9$He) can be described at low energies as a $^8{\rm C}+p(n)$ two-body system. The effective Hamiltonian used for this study includes a core-valence potential with nuclear and Coulomb parts, which have been taken in the Woods-Saxon (WS) and dilatation-analytic forms\,\cite{Wang2017}, respectively. The WS potential parameters are: depth $-64$\,MeV, spin-orbit strength 15\,MeV, diffuseness 0.7\,fm, and  radius 2.2\,fm. 
The predicted three lowest Gamow poles of $^9$N in our model are: $\tilde{E}(1/2^+)=(1.14-i1.54)$\,MeV, 
$\tilde{E}(1/2^-)=(1.99-i0.39)$\,MeV,
and $\tilde{E}(5/2^+)=(5.31-i1.00)$\,MeV.

To study the long-time behavior of two-nucleon  decay, we focus on the lightest $2p$ emitter $^6$Be, whose $2p$ decay energy is measured to be $Q_{2p}=1.372$\,MeV with the width $\Gamma=92$\,keV \cite{ENSDF}. In our framework, the system is  viewed as an  $\alpha+p+p$ three-body system. The corresponding configurations in the Jacobi coordinates are labeled by  quantum numbers  $(K,\ell_x,\ell_y)$, where $\ell_x$ is the orbital angular momentum of the neutron pair with respect to their center of mass,  and $\ell_y$ is the pair's orbital angular momentum with respect to the core. The Pauli-forbidden states occupied by the core nucleons are eliminated
according to Ref.\,\cite{Sparenberg1997}. The valence-nucleon interaction  is represented by the finite-range Minnesota force with the  parameters of Ref.\,\cite{Thompson1977} except for the exchange-mixture coefficient $u$ = 0.9. The effective $\alpha$-$n$ interaction is described by a WS potential with parameters from Ref.\,\cite{Wang2017} except for  depth $-49$\,MeV. The obtained ground state of $^6$Be has a complex energy $(1.39-i0.04)$\,MeV.  

For the two-body initial states of $^9$N and $^9$He, the Berggren basis is adopted with a deformed scattering contour in the complex momentum plane, along the path: $k=0 \rightarrow -0.2-i0.3 \rightarrow 0.3-i0.3 \rightarrow 0.6 \rightarrow 6.0$ (in fm$^{-1}$). Each segment was discretized with 70 scattering states. The initial state of $^6$Be is calculated in the three-body framework. To investigate the universal property of three-body decay, we have also constructed an artificial two-neutron ($2n$) emitter $^6$He$^\prime$ with identical parameters except for the WS depth $V_0$ = $-43$\,MeV. The three-body calculations of $^6$Be and $^6$He$^\prime$ were carried out in a model space defined by $\max(\ell_{x}, \ell_{y})\le 7$ and for a maximal hyperspherical quantum number $K_{\rm max} = 20$. In the hyperradial part, we used the Berggren basis for the $K \le 7$ channels and the harmonic oscillator basis with the oscillator length of 1.75\,fm and $N_{\rm max} = 20$ for the remaining channels. For the GCC calculation of the initial state, the complex-momentum scattering contour is given  by the path $k = 0 \rightarrow 0.2-i0.05 \rightarrow 0.4  \rightarrow 1.0 \rightarrow 6.0$ (in fm$^{-1}$), and discretized with 70 scattering states for each segment.

To study time evolution, the initial complex-energy Gamow state  is decomposed into real-momentum scattering states using the Fourier-Bessel series expansion in the real-energy Hilbert space \cite{Baz1969} and then propagated with the corresponding real-momentum contour. Each segment is discretized with 140 scattering states, and 420 more are added in the energy interval [$E_r-2\Gamma$, $E_r+2\Gamma$] to increase the precision. The spectral function $\rho$ is obtained with the wave function $\Psi(t)$ taken at a long time $t$ = 40$T_{1/2}$.  
Since the Coulomb potential and kinetic energy do not commute in the asymptotic region, there is analytical solution for the charged three-body system. Therefore, for $^6$Be, we only consider the interactions inside the sphere of radius 400\,fm, but the wave function is still defined in the momentum space beyond this cutoff.

\section{Scenarios to probe non-exponential decay}

\subsection{Non-exponential decay of a  threshold resonance}

\begin{figure}[htb]
\includegraphics[width=1.0\columnwidth]{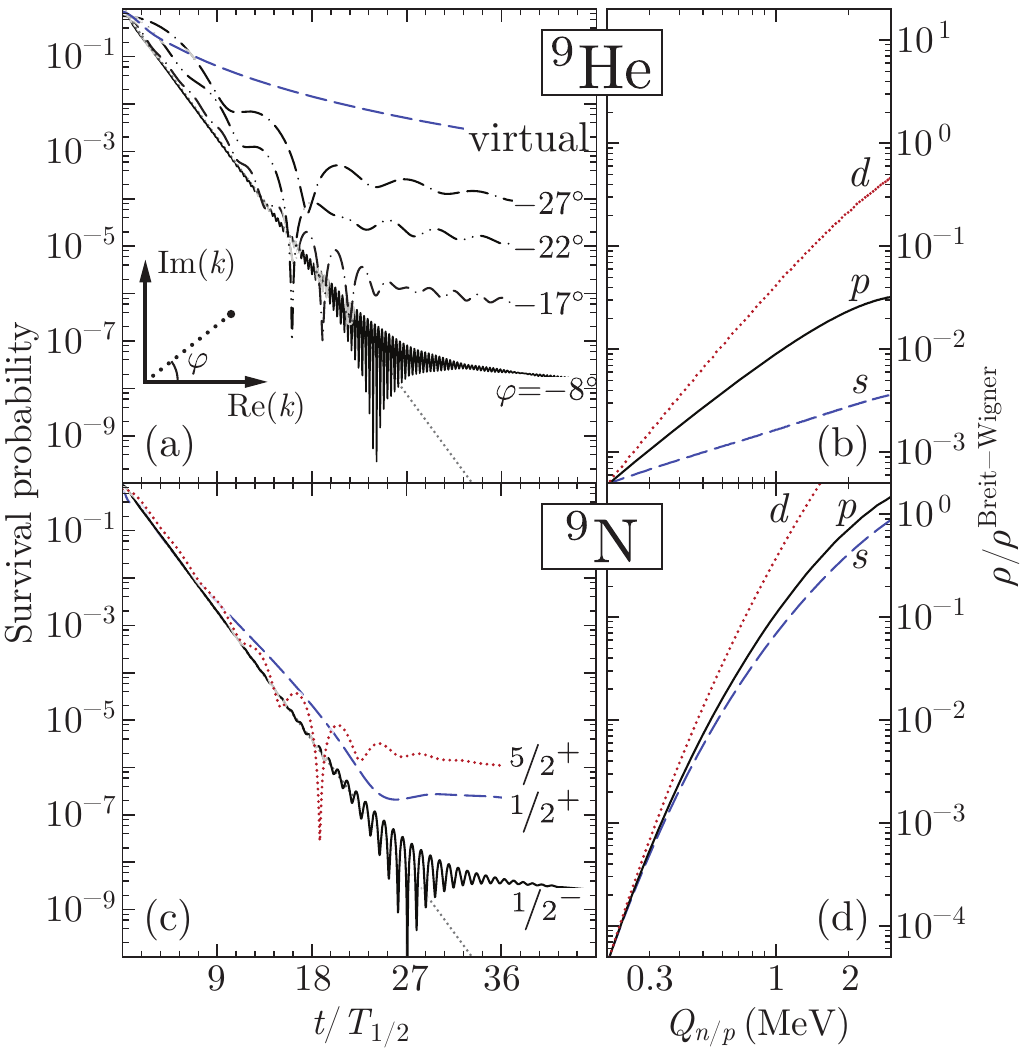}
\caption{Survival probability $\mathcal{S}(t)$ as a function of time (relative to $T_{1/2}$) for (a) the 1/2$^-$ state of $^9$He for different depths $V_0$ of the WS potential, and (c) the low-lying states of $^9$N.
The near-threshold behavior of the spectral function $\rho$ (relative to the Breit-Wigner distribution) is shown in (b) for  neutron and in (d) for  proton $s,p, d$ partial waves. The polar angle $\varphi$  indicates the location of the resonant state in the complex-$k$ plane. Also shown is the survival probability for the virtual 1/2$^+$ state
in $^9$He. For this state,  $T_{1/2}$ was assumed to be 20\,fm/$c$. }\label{Survival_probability}
\end{figure}

The survival probability of decaying state depends on the energy distribution of   the underlying spectral function. For a  system that
decays  exponentially, one expects a Breit-Wigner type distribution. However, this does not hold for near-threshold states having large decay widths\,\cite{Khalfin1958,Baz1969,Greenland1988,Esposito2004,Miyamoto2006}. 
In fact, the time evolution of any resonance involves exponential  and non-exponential components\,\cite{Ramrez2019,Ramrez2021}. Since exponential components decay faster, a transition to a power-law regime is bound to take place eventually  (see Fig.\,\ref{Survival_probability}a,c). 

While this behavior is universal, the actual dynamics is determined by the structure of the initial state, decay channel, and most of all, the nature of the scattering continuum that drives the post-exponential decay. To illustrate this concept, we have analyzed the survival probability of the 1/2$^-$ resonant state in $^9$He by varying the depth of the WS potential. The obtained resonant states are identified by  the polar angle $\varphi = -\cot^{-1}(2E/\Gamma)/2$, which reveals their location in the complex-$k$ plane and provides an estimation of the non-exponential component. As seen in Fig.\,\ref{Survival_probability}a, the deviation from the exponential decay quickly increases as $\varphi$ moves towards $-45^{\circ}$. 
This agrees with the finding of earlier studies\,\cite{Greenland1988,Rothe2006,Mercouris1997} suggesting that the post-exponential decay is expected to take over rather quickly -- hence easier to be observed -- in  threshold resonances with $E_r\approx\Gamma$. 

In this sense, unbound nuclides, such as the proton-rich $^9$N, could be the perfect candidates for observing the non-exponential decay. Similar to 
$^9$He, the predicted 1/2$^+$ and 5/2$^+$ states of $^9$N with relatively large decay widths are expected to transition to the power-law decay earlier than the 1/2$^-$ state (see Fig.\,\ref{Survival_probability}c) as their structure is strongly influenced by the scattering continuum. Moreover, as the analog of the $s$-wave virtual state\,\cite{Ohanian1974,Wang2019}, the 1/2$^+$ state is dominated by  non-resonant components. Consequently, for this state, the transition from exponential to non-exponential decay is fairly gradual.

\begin{figure}[htb]
\includegraphics[width=0.9\columnwidth]{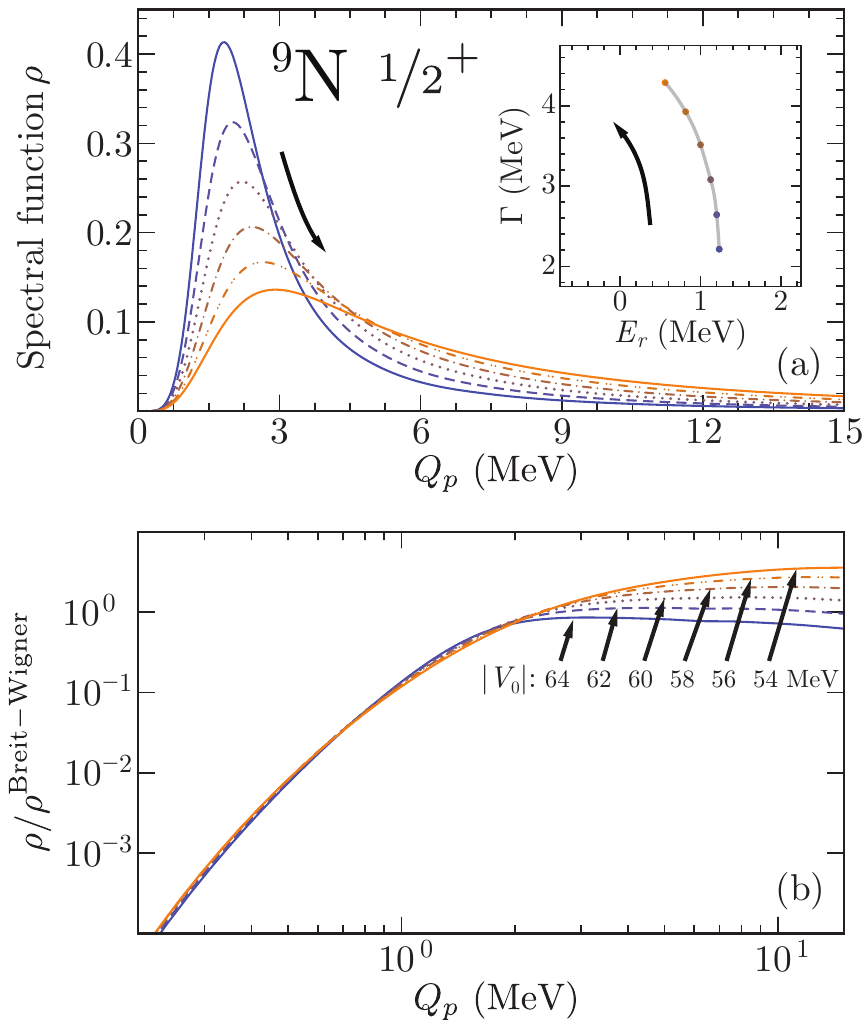}
\caption{(a) Predicted spectral functions for the ground state of $^9$N versus decay energy for different WS depths $|V_0|$. The arrow shows the direction of evolution of the spectral function and complex energy (in the insert) as $|V_0|$ decreases from 64 to 54\,MeV. (b) The near-threshold behavior of the spectral functions relative to the Breit-Wigner distribution.}\label{Line_shape}
\end{figure}

The channel-$\ell$ dependence of the non-exponential decay can be related to the Wigner cusp phenomenon\,\cite{Wigner1948}, and it manifests itself in different threshold behavior of the spectral function. Generally, the spectral function $\rho$ is governed by the centrifugal barrier and the Coulomb interaction near the threshold, and it approaches  the Breit-Wigner  distribution as the energy increases (see Fig.\,\ref{Line_shape}). For the neutrons, the spectral functions corresponding to states
with different  $\ell$-values approach the low-energy scattering limit quite differently (see Fig.\,\ref{Survival_probability}b), which results in a power law
behavior $\mathcal{S} \propto 1/t^{2\ell+3}$  at long times\,\cite{Torrontegui2010,Fonda1978,Peshkin2014}. For protons, the Coulomb interaction greatly reduces  the $\ell$-dependence at low energies, see Fig.\,\ref{Survival_probability}d. 

Formally, the onset of  non-exponential decay at long times is determined by the low-energy scattering  impacting the near-threshold properties of the spectral function. The two most important factors here are the behavior of the density of states in the particular channel  and the structural properties of the system. 
While in dripline nuclei the continuum appears at very low
energies, their production rates are usually low, and this makes
it more suitable for the experiment with high statistics  to observe the transition  directly in $\mathcal{S}(t)$. 

\subsection{Two-particle decay observables}

The two-nucleon ground-state decay is a rare process observed  in a handful of nuclides\,\cite{Pfutzner2012,Blank2008,Pfutzner2023}. 
In contrast to a two-body final state, in the three-body case the conservation of energy, momentum, and angular momentum  is not
sufficient to uniquely identify the asymptotic state. This allows for configuration mixing and competition between different
intrinsic configurations \cite{Wang2019,Wang2022}. This structural information is seen in  the asymptotic nucleon-nucleon correlations that can be directly measured\,\cite{Miernik2007,Pfutzner2012}.
The presence of the non-exponential component in $\mathcal{A}(t)$ is expected to affect    nucleon-nucleon correlations, thus providing a unique window into the  decay process.

\begin{figure}[htb]
\includegraphics[width=1.0\columnwidth]{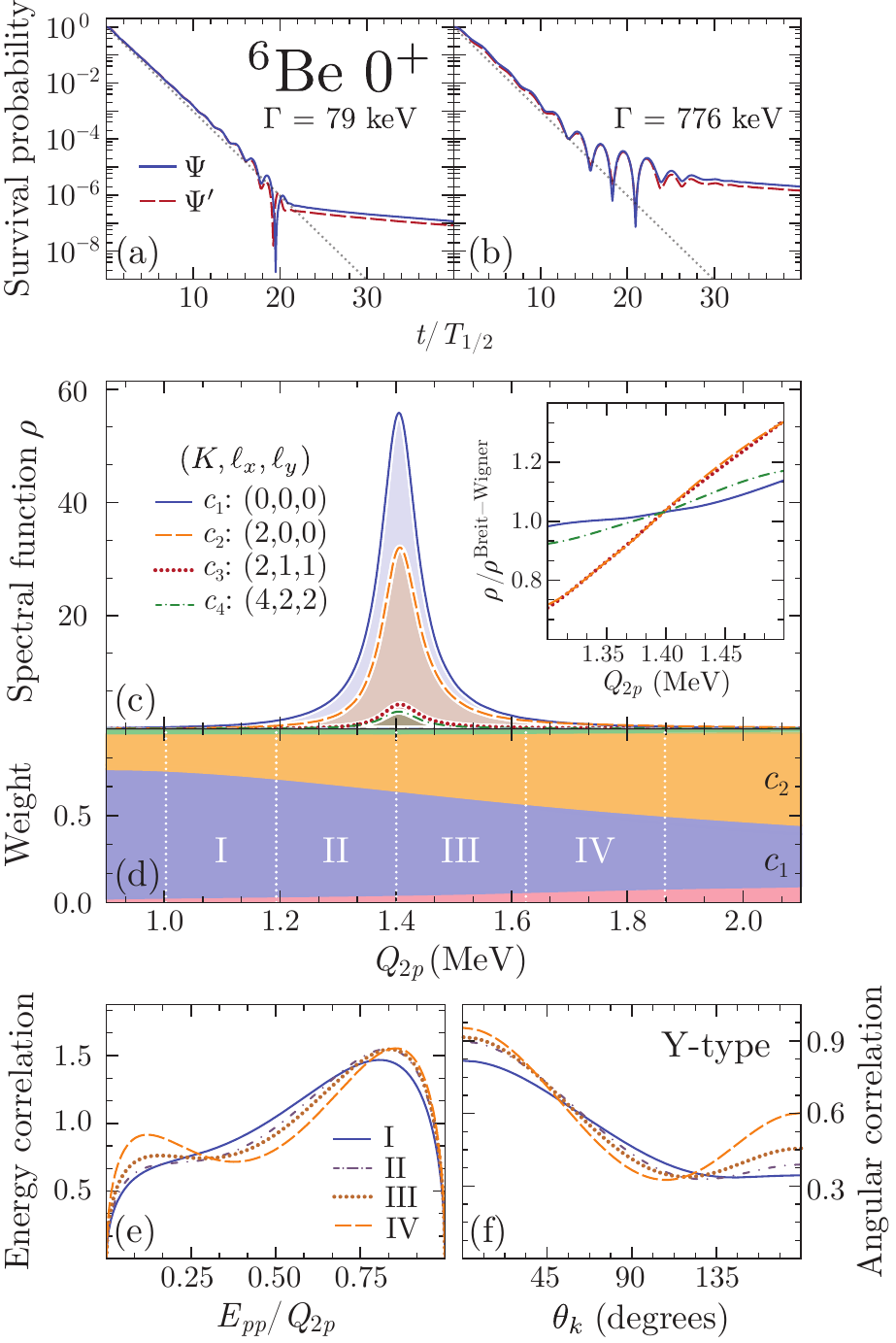} 
\caption{Survival probability of the $^6$Be ground-state $2p$ decay, in which $\Psi$ (solid lines) and $\Psi^\prime$ (dashed lines) represent the initial wave functions obtained by the expansion of the Gamow state in the real-momentum Fourier-Bessel basis and harmonic-oscillator basis, respectively. Two cases are considered: (a) Ground state with $\Gamma=79$\,keV and (b) broad resonance with $\Gamma=776$\,keV.  (c) Calculated spectral functions of the different configuration components $c_i$ of the $^6$Be ground state;  the insert shows the ratio between the predicted energy distribution and the Breit-Wigner form factor. (d) The  weights of $c_i$ as a function of the $2p$ decay energy $Q_{2p}$. Asymptotic energy (e) and  angular (f)  correlations for different  energy ranges  in Jacobi-T (left) and Jacobi-Y (right) coordinates.}\label{Configuration_distribution}
\end{figure}

 To demonstrate the  post-exponential characteristics of two-nucleon decay, we consider the case of $2p$ decay of $^6$Be. 
 The precise form of the initial state is likely to be influenced by the specific production mechanism, which may not necessarily result 
 in a Gamow state. 
To assess the associated uncertainty, in addition to the real-momentum projected Gamow state $\Psi(0)$, we consider an   initial state $\Psi^\prime(0)$ generated by 
the harmonic-oscillator expansion of $\tilde{\Psi}(0)$.
Both yield practically identical results for the survival probability even for a broad resonance (see Fig.\,\ref{Configuration_distribution}a,b). This is due to the fact that, once a resonance is formed, the main component of the wave function inside and around the nucleus is almost fixed.

For $^6$Be, the transition from exponential to power-law decay takes place at $t\approx$ 20$T_{1/2}$. This indicates that -- in order to extract information about the post-exponential decay directly from the spectral function -- one requires an energy resolution that is much finer than the resonance width, as dictated by the uncertainty principle. Nevertheless, it doesn’t necessarily impact other physical observables, especially for those with accumulated effects. As shown in Fig.\,\ref{Configuration_distribution}d, the ground state of this three-body system is composed of multiple Jacobi-coordinate configurations, whose relative weights in the spectral function undergo significant variations as a function of the decay energy. Furthermore, although suppressed by the Coulomb interaction, each individual configuration $c_i$ possesses a distinct spectral function that deviates from the Breit-Wigner shape (see Fig.\,\ref{Configuration_distribution}c). This can be attributed to
the hyperspherical quantum number $K$, which represents the centrifugal barrier for three-body decays\,\cite{Raynal1970,Wang2017}.

\begin{figure}
\includegraphics[width=0.9\columnwidth]{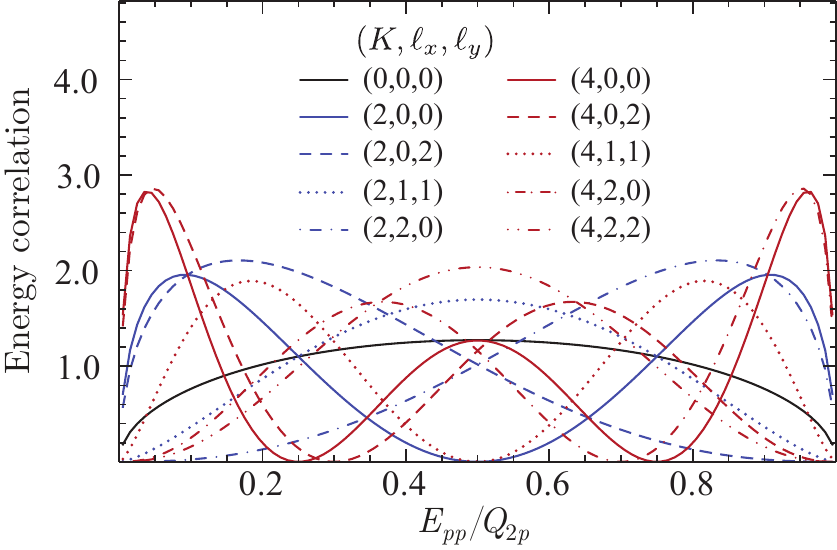}
\caption{Nucleon-nucleon correlations $\braket{c_i |\delta(\epsilon - E_{pp}/Q_{2p}) | c_i}$ in the  individual Jacobi configurations $c_i$'s. $E_{pp}$ is the relative kinetic energy between the emitted neutrons, and $Q_{2p}$ is the decay energy.}\label{Configuration_correlation}
\end{figure}

Due to the factor that the long-time behavior is determined by the asymptotic three-body configuration $(K,\ell_x,\ell_y)$ distinctively (see Fig.\,\ref{Configuration_correlation}), these cumulative changes among configuration components leave an imprint on the  asymptotic correlation of emitted nucleons, which results in appreciably different patterns in different regions of $Q_{2p}$ as shown in Fig.\,\ref{Configuration_distribution}e,f.
Therefore, by binning the resonance peak at different energies, one should be able to assess the $Q_{2p}$-evolution of the energy- and angular correlations. This could provide useful information on nuclear structure and provide an indirect evidence for the transition to the non-exponential decay regime. This is supported by Ref.
\cite{Egorova2012}, where the $2p$ energy correlation 
in the $^6$Be decay becomes more pronounced for both small and large $E_{pp}$/$Q_{2p}$  as $Q_{2p}$ increases, consistent with our predictions.

\begin{figure}[htb]
\includegraphics[width=1\columnwidth]{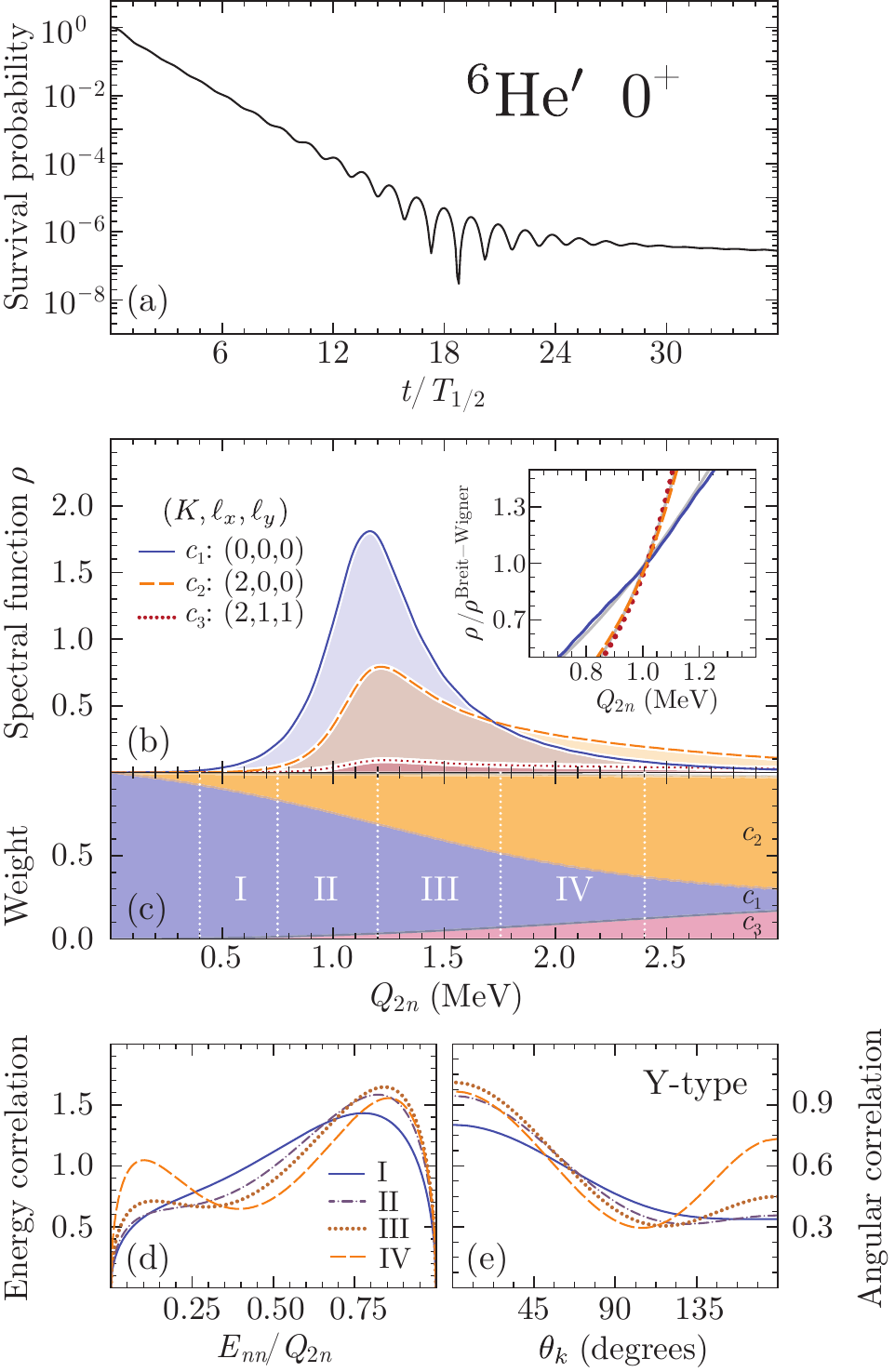}
\caption{(a) Survival probability of the $^6$He$^\prime$ $2n$ decay, the ground state of $^6$He$^\prime$ has a complex energy $(1.28-i0.15)$\,MeV. (b) Calculated spectral functions of the different configuration components $c_i$ of the $^6$He$^\prime$ ground state;  the insert shows the ratio between the predicted energy distribution and the Breit-Wigner form factor, which is approximately proportional to $Q_{2n}^{K+2}$ (solid gray lines) in the plotted range of $Q_{2n}$. (c) The weights of $c_i$ as a function of $Q_{2n}$. Asymptotic energy (d) and  angular (e)  correlations for different  energy ranges  in Jacobi-T (left) and Jacobi-Y (right) coordinates.}\label{Configuration_distribution_He6}
\end{figure}

The energy dependence of asymptotic correlations and the non-exponential behavior are universal in three-body decays, as long as the decaying structure is not too narrow or dominated by a single configuration. This is  satisfied for  two-nucleon emitters in light nuclei. 
In particular, as illustrated by an example of  $^6$He$^\prime$ shown in Fig.\,\ref{Configuration_distribution_He6}, 
similar post-exponential behavior is expected to be present in the $2n$ decay. In this case, the energy dependence of particle correlations is enhanced in the absence of the Coulomb interaction.

\subsection{Interference between near-lying states}

\begin{figure*}[htb]
\floatbox[{\capbeside\thisfloatsetup{capbesideposition={right,top},capbesidewidth=4cm}}]{figure}[\FBwidth]
{\caption{Interference between two close-lying  $0^+$ resonances in $^6$He$^\prime$ for the three values of the energy splitting of the doublet $\Delta E$ (in MeV). Left: Spectral functions versus decay energy. The arrow indicates the suppression of the spectral function of $|2\rangle$. Right: The time dependence of the corresponding survival probabilities. The decay widths (in keV) of the doublet ($\Gamma_1$, $\Gamma_2$) are (34, 60), (30, 52), and (6, 68) for large, moderate, and small values of $\Delta E$, respectively.}\label{Two_level_system_full}}
{\includegraphics[width=0.75\textwidth]{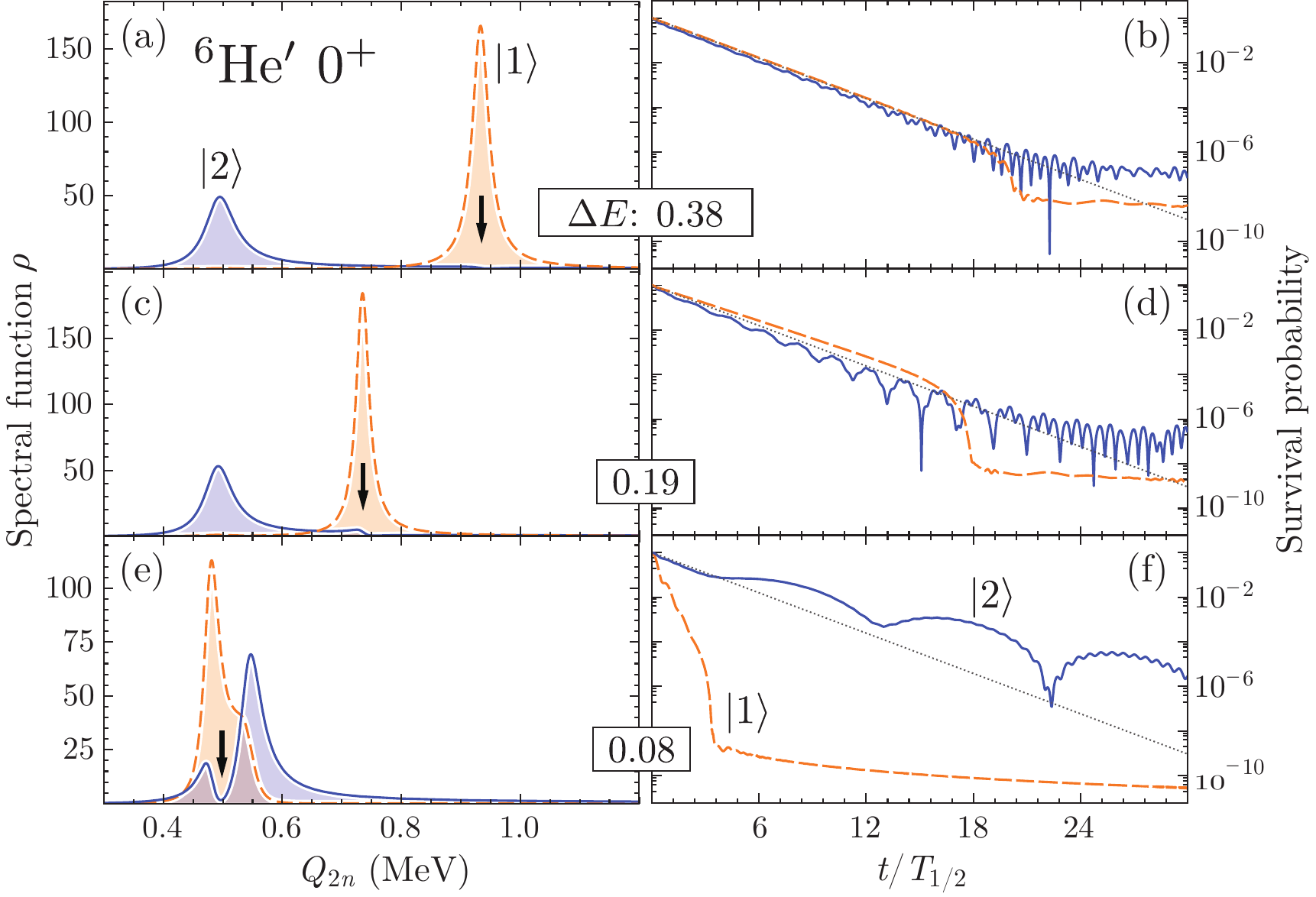}}
\end{figure*}

Threshold effects  may impact resonance structures, especially when two or more states with the same spin-parity lie close in energy\,\cite{Onley1992,Peskin1994,Delamadrid2017}. In this case, the overlapped resonances rearrange the decay widths by increasing coupling with the continuum \,\cite{Sokolov1988,Volya2003,Magunov2003,Kravvaris2017}. As a result, the decay width of one of the resonances  becomes collectively enhanced. To illustrate this point and to show the effect of the continuum coupling on  the  spectral functions of  the overlapping resonances, we consider a two-level 0$^+$ system in $^6$He$^{\prime}$ by readjusting the depth  of the WS potential.

The excited state $|1\rangle$ is dominated by the $d^2$ configuration while the  ground state $|2\rangle$ is primarily $p^2$. Figure\,\ref{Two_level_system_full} shows the evolution of the spectral functions and corresponding survival probabilities for different energy splittings $\Delta E=|E_r(1)-E_r(2)|$ of the doublet. When $\Delta E$  is  large, there is only a small suppression at the tail of spectral function of $|2\rangle$ and both states are characterized by comparable widths.  As the states begin to overlap, a strong interference occurs that significantly impacts the spectral functions of the doublet (see Fig.\,\ref{Two_level_system_full}e,f). 
The corresponding survival probabilities show dramatic deviations from  the exponential decay regime. In particular, the state $|1\rangle$  decays much faster than suggested by its intrinsic decay width, whilst the state $|2\rangle$  exhibits a remarkably slow decay, which is in accord with the discussion of Refs.\,\cite{Volya2003,Kravvaris2017}. 

\begin{figure}
\includegraphics[width=0.9\columnwidth]{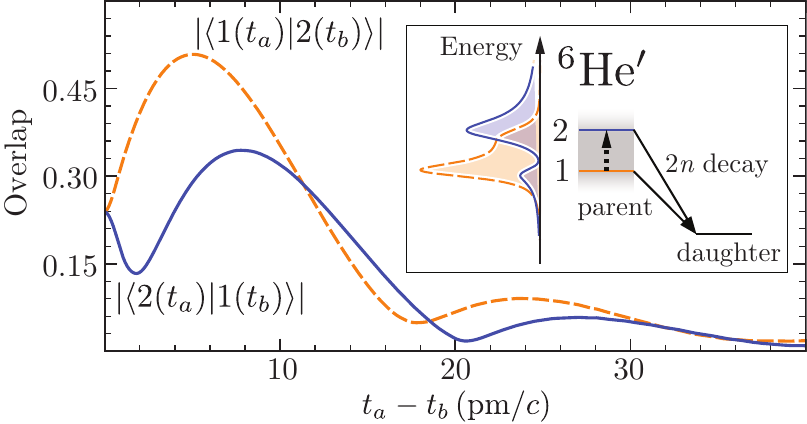}
\caption{Transition rate (overlap) between one state of the doublet (at time $t_a$) and the second state (at time $t_b$) in $^6$He$^\prime$.  The decay diagram is shown in the inset. The virtual transition between the doublet states is marked by the arrow.}\label{Two_level_transition}
\end{figure}

Such exponentiality during the decay process could occur between any near-lying resonances of the same spin-parity, due to the virtual transition governed by  the scattering continuum and the difference between the orbital angular momentum structure of the doublet states. Assuming that the initial wave packet is a mixture of $| 1 \rangle$ and $| 2 \rangle$,  one notices that  the survival amplitude (\ref{Survival_probability_expression}) impacted by the overlaps of the near-lying resonances at different times, namely $\braket{1(0) | 2( t)}$ and $\braket{2(0) | 1( t)}$. Due to the Hermitian property of the time evolution operator, one obtains:
\begin{equation}\label{Overlap}
     \braket{1(\delta t) | 2(0)} = \braket{1(0) | 1(-\delta t)} = \braket{1(t_a) | 2(t_b)},
\end{equation}
where $\delta t = t_a - t_b$. This means that, while the wave function undergo an overall exponential decay, their remaining components are ``translationally'' invariant with respect to the choice of the initial time. Hence, the intrinsic changes in the resonant wave function depend on the difference  $t_a - t_b$ . 
In the case of $^6$He$^\prime$, the overlaps of the two unbound $0^+$ states have been calculated and shown in Fig.\,\ref{Two_level_transition}. 
When $t_a > t_b$, $|\braket{1(t_a) | 2(t_b)}|$ is more likely to be larger than $|\braket{2(t_a) | 1(t_b)}|$. It indicates that time-delayed ($t_a - t_b > 0$) transition is more likely to occur from $| 2 \rangle$ to $| 1 \rangle$ while the time-prompt transition $ (t_a - t_b < 0)$  ---  from $|1\rangle$ to $| 2 \rangle$.

\begin{table*}
\caption{Different scenarios to access non-exponential decay and their characteristics 
(advantages {\bf A} and requirements {\bf R}).
The  promising candidates for experimental tests are listed.}\label{Requirements_Candidates}
\begin{ruledtabular}
\begin{tabular}{l>{\raggedright}p{10cm}>{\raggedright\arraybackslash}p{4cm}}

Scenario & Characteristics & Candidates \\
\\[-6pt] \hline 
\\[-6pt]
Threshold resonance
& 
{\bf A}: Pronounced non-exponentiality;
very short half-life. 
{\break}{\bf R}: Low partial waves involved;
sufficient statistics
& 
$^9$N\,\cite{Charity2022}, $^9$He\,\cite{AlKalanee2013,ENSDF} 
\\ 
\\[-6pt] \hline 
\\[-6pt]
 Three-body decay
& 
{\bf A}: Nucleon-nucleon correlations accessible;
accumulated energy dependence. {\break}{\bf R}:
Appreciable decay width allowing for energy binning;
configuration mixing involved;
high statistics required
& 
$^6$Be\,\cite{Egorova2012}, $^{13}$Li\,\cite{Andre2022}, $^{16}$Be\,\cite{MonteagudoGodoy2019} 
\\
\\[-6pt] \hline 
\\[-6pt]
Near-lying resonances
& 
{\bf A}: Pronounced non-exponentiality;
distorted spectral function due to interference effects.
{\break}{\bf R}: Doublets of states with identical $J^\pi$ quantum numbers; high statistics required
&  $^{13}$C\,\cite{Kravvaris2017,ENSDF}, $^{13}$N\,\cite{Hanselman2022} 
\end{tabular}
\end{ruledtabular} 
\end{table*}

Although the system discussed here is artificial, experimental candidates for   doublet states with appreciably different decay widths have been suggested in Ref.~\,\cite{Kravvaris2017}. In particular, the two near-lying 3/2$^+$ resonances in $^{13}$N have been generated simultaneously through the proton transfer reaction $^{12}$C($^3$He, $d$)$^{13}$N$^*$\,\cite{Hanselman2022}, and the twisted spectral functions have been observed.  In this scenario, the non-exponential decay can be  characterized by directly observing the decay pattern or through the analysis of the spectral function. We note that the presence of the  interference  dramatically reduces the need for a stringent energy resolution.

\section{Summary}

A wave function collapse onto a stationary state is one of the fundamental principles of quantum mechanics.  Resonances and their long-term features studied in this work offer a remarkable intermediate perspective on a non-destructive collapse where part of the wave function carries initial
information.
The phenomenon of a  non-exponential decay of an open  quantum system at long times   can be traced back to spectral function  exhibiting characteristic threshold behavior. 
Consequently, decays of near-threshold states/structures are expected to directly probe the non-exponential decay regime. 

The wealth of the current results on two-particle decays of rare isotopes and studies of  overlapping resonances offer new opportunities for investigations  of the 
non-exponential component. The decay  scenarios discussed in this paper are summarized in Table\,\ref{Requirements_Candidates}.
In particular, we demonstrate that useful insights into survival amplitude ${\cal A}(t)$ can be offered by analyzing the energy dependence of the asymptotic particle-particle correlations. From such data, one could obtain structural  information about the decaying state that would enable one to conclude whether the  non-exponential decay phase has indeed been reached.  Another phenomenon with possible experimental consequences  is the interference of near-lying decaying states of the same  quantum numbers.  In this case, the states may decay non-exponentially through a virtual transition. 
 
 Admittedly,  the direct observation of the non-exponential component of ${\cal S}(t)$ is going to be  difficult. Still, we believe that  the scenarios proposed in this work can offer an indirect evidence for this fundamental property of open quantum systems and offer new avenues for future explorations.

\section*{Acknowledgements}

This material is based upon work supported by the National Key Research and Development Program (MOST 2022YFA1602303); the National Natural Science Foundation of China under Contract No.\,12147101; the U.S.\ Department of Energy, Office of Science, Office of Nuclear Physics under award numbers DE-SC0013365 (Michigan State University), DE-SC0009883 (Florida State University), and DE-SC0023175 (NUCLEI SciDAC-5 collaboration).

\bibliography{references}
\end{document}